\documentclass[fleqn,twoside]{article}
\usepackage{espcrc2, epsfig}

\newcommand{\AmS}{{\protect\the\textfont2
  A\kern-.1667em\lower.5ex\hbox{M}\kern-.125emS}}

\title{{\vskip-20pt \hfill {\rm\normalsize UCLA/03/TEP/27}}\break
Dark matter searches: looking for the cake or its frosting?
\break \hfill
(Detectability of a subdominant component of the CDM)}

\author{Graciela Gelmini\address[UCLA]{Department of Physics and Astronomy,
 UCLA, Los Angeles, CA 90095-1547, 
USA}\thanks{gelmini@physics.ucla.edu}\thanks{Talk presented at TAUP 2003,
Seattle, USA, 5-9 Sept. 2003, based on the papers 
in reference \cite{duda}.}\thanks{Work supported in part by
the U.S.\ Department of Energy Grant
No.\ DE-FG03-91ER40662, Task C.}}

\begin{document}

\begin{abstract}
The dark matter candidates we are searching for,
e.g.  neutralinos, may be
 one of many components of the cold dark matter (CDM).
We point out here that very subdominant components, constituting
even 1\%  of the CDM for indirect detection and much less for direct
detection, remain observable in current and future searches.
 So if a CDM signal is confirmed in CDM search experiments 
(except for a signal from annihilations in
the dark halo) we will need to find out the halo fraction accounted for
the CDM component we detected.

\vspace{1pc}
\end{abstract}

\maketitle

%\section{FORMAT}

We live in a Universe with a complex composition, about 
0.7 of the total density is in dark energy and 0.3 of it in cold dark matter (CDM),
with smaller components of baryons and neutrinos. This is a composition we
would not have guessed fifteen years ago. It is thus conceivable that the
composition of the CDM could be as complex, the
 Weakly Interacting Massive Particle (WIMP) candidate we are
looking for in direct and indirect searches being only one of the components.
 The most common WIMP candidate is the neutralino (although there are other
 possible lightest supersymmetric particle candidates),
We find many other possible DM candidates in the literature such as
 axions,  Q-balls, heavy non-thermal relics,  warm or cool DM,
 interacting DM, etc. which cannot be found in WIMP searches.

So the questions that come to mind are: if  WIMPs
constitute only a fraction, $f$, say 1\% or less, of the local dark matter halo,
would these particles still be observable in the current and proposed direct
and indirect dark matter searches?  If we confirm a CDM signal in any of
our searches, have we found the bulk of the CDM? The answers to these
questions, found in the literature since the inception of the subject 
(e.g.~\cite{duda,bottino2341} and references therein), are:
yes! to the first and no! to the second. However
in \cite{duda} we explored for the first time  how small the 
 halo fraction $f$ can
be for detectable neutralinos and additionally, if  a
combination of direct and indirect detection data  could tell the value of $f$.

Although the event rate $R$ in direct detection searches is proportional to
the local WIMP density $\rho_{\chi}$, it does not necessarily decrease linearly
with $\rho_{\chi}$, as naively could be expected. This is because the local
WIMP density is proportional to the cosmological WIMP density $\Omega_{\chi}$,
which in turn (for all  particles that have been in equilibrium in the
early Universe) is inversely proportional to the  WIMP annihilation cross section
(in the early Universe), $\Omega_{\chi} \sim \sigma_a^{-1}$. Thus, a
reduction in $\rho_{\chi}$ requires an increase in $\sigma_a$. This increase is often
associated with an increase in the scattering cross section $\sigma_s$ of WIMPs
off atomic nuclei. In fact there are ``crossing symmetries'' that relate
the  amplitudes of both processes and both increase with decreasing mediators
mass and increasing coupling constants. The relation between $\sigma_a$ and
$\sigma_s$  is complicated in many ways, however in many instances it holds
that they increase in the same manner.  Then, in  direct detection
rates the decrease in density is
compensated by an increase in $\sigma_s$  and 
 rates may remain high even for very subdominant WIMPs,
$ R \sim \sigma_s \rho_{\chi}
 \sim \sigma_s\sigma_a^{-1} \simeq constant$. 

In  \cite{duda}  models implemented in the ``DarkSUSY" code were used in
which  a neutralino
is the CDM candidate, and the halo and cosmological neutralino 
fractions were taken to be the same, 
$f = \Omega_{\chi}/\Omega_{DM}= \Omega_{\chi} h^2/ 0.15$
 ($\Omega_{\chi} =0.3$ and $h=0.7$). The latter is correct if  
segregation mechanisms for different CDM components are not important, e.g.
if all  components are non-collisional. 
{\epsfysize 7cm \epsfbox{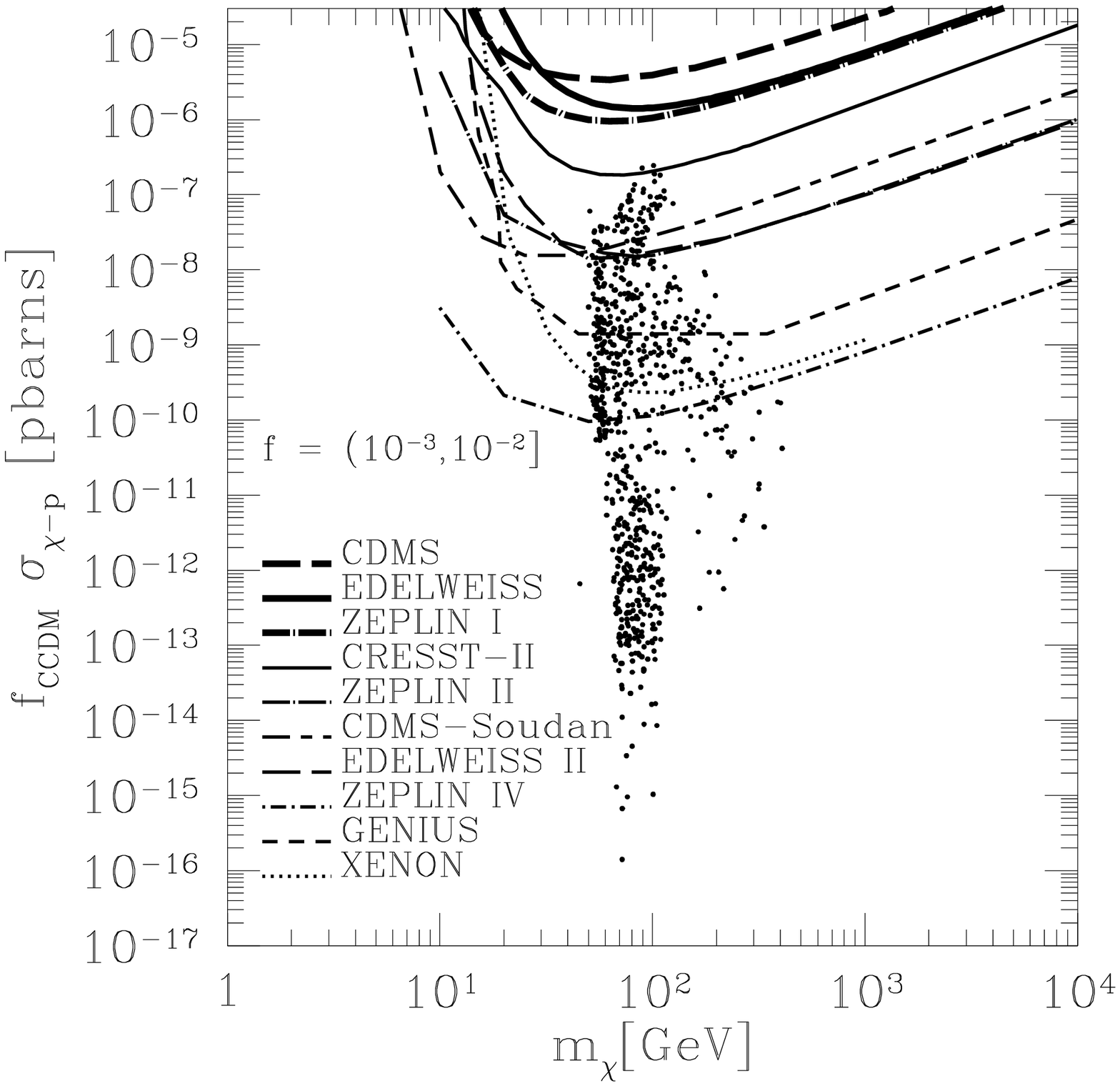}} 
Figure 1. Each point corresponds to a  neutralino
 model with $f$ between 0.1 and 1\% together  with present direct detections bounds
 (thicker lines) and future direct detection discovery limits.
\vspace{0.2cm}

The DAMA region was not explored in \cite{duda} 
 (the models used provide practically
no candidates in that region), however subdominant candidates, even
 with halo fraction 1\%
or  smaller, are among the candidates in this region
studied by the Torino group \cite{Torino-DAMA}, for example. If the DAMA
signal is not confirmed, Fig.~1 shows that CRESST-II could probe neutralino
halo factions $f > 10^{-3}$. In \cite{duda} is shown that  
ZEPLIN II, CDMS at the Soudan mine and EDELWEISS II could probe  $f > 10^{-4}$.
 These are experiments from which data are expected in the next few
years.  Further in the future, another generation of detectors, such as ZEPLIN
IV, CryoArray, GENIUS and XENON, may be able to reach down to neutralino
 halo fractions of
order $10^{-5}$, which are the smallest found  in \cite{duda}.

Concerning indirect detection, the flux of rare
cosmic rays and of gamma-rays produced in halo annihilations depend on the
product of the square of the density and the annihilation cross section into a
particular channel, $\sigma_a {\rho_\chi}^2$. Thus, even if an increase
in the cross section compensates for the decrease in one of the powers of the
density, the fluxes  still decrease linearly with the  halo WIMP density.
 Hence WIMPs which are a
subdominant component of the total CDM would be extremely difficult to detect
through this method, as already concluded in \cite{duda,bottino2341}
(e.g. see Fig.~2).
{\epsfysize 7cm \epsfbox{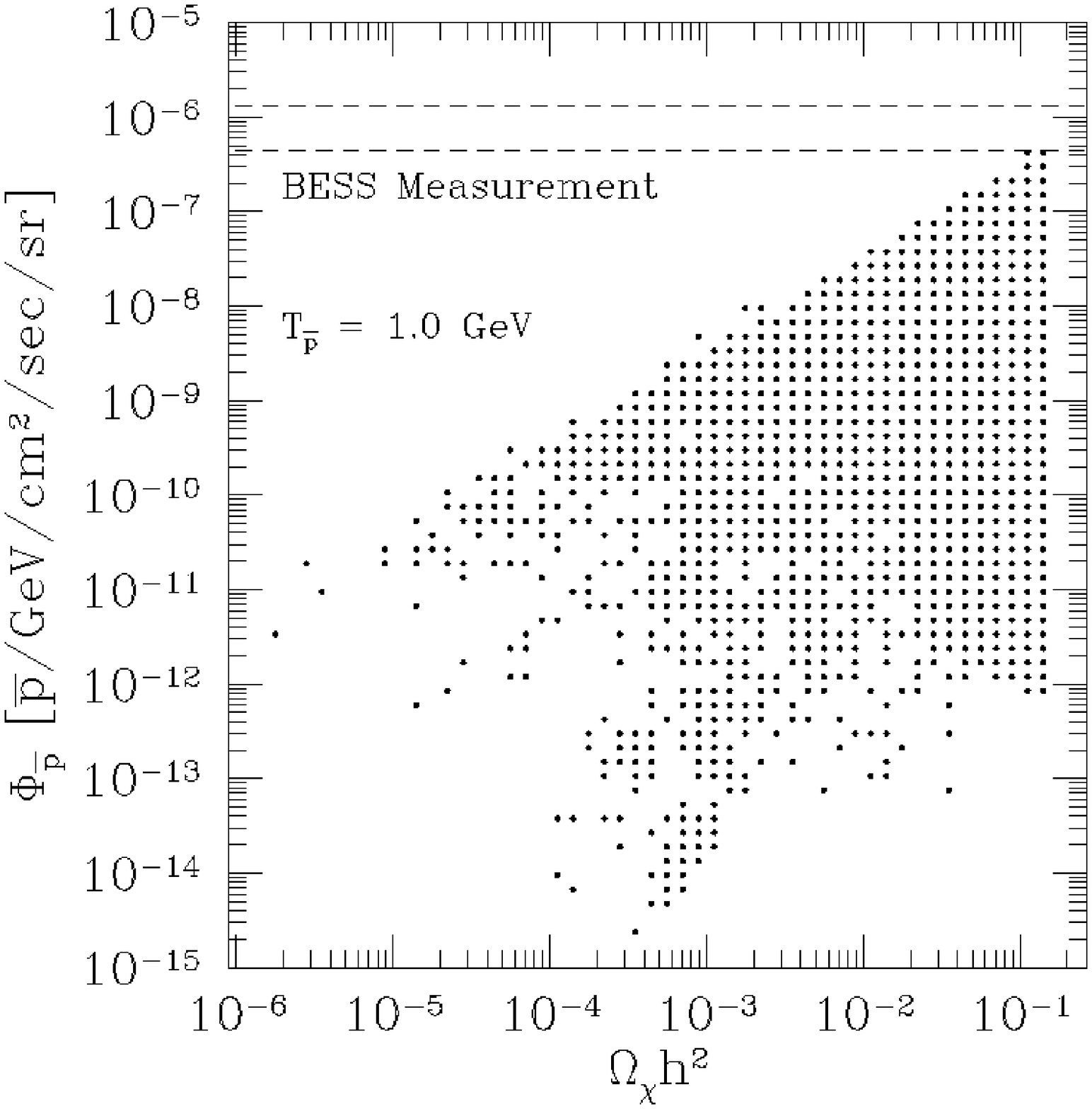}}
Figure 2.  Solar modulated anti-proton 
flux at 1 GeV kinetic energy at solar
  minimum (using the force-field approximation in DarkSUSY) as a
  function of the neutralino relic density.  A regular grid of points shows the
  region  with models.
\vspace{0.2cm}

The intensity of the high-energy neutrino emission from neutralino 
annihilations in the Sun
and the Earth, instead,  would in many cases remain high with decreasing density.
As the Sun and Earth sweep through
the dark matter halo interactions with nuclei within these bodies slow
WIMPs enough so they can become gravitationally captured. It is
the capture rate $C$ which becomes constant when the ``compensation
argument'' holds, $C \sim \sigma_s \rho_{\chi} \simeq constant$.
At small times $t$ the number of capture neutralinos  
$N_\chi \simeq Ct$, increases with time $t$ and also the annihilation rate increases
 $\Gamma_a \sim \sigma_a N_\chi^2\sim \sigma_a C^2 \sim \sigma_a
\sim \Omega_\chi^{-1}$ thus we could expect ``overcompensation'', namely an
increase in the rate for subdominant WIMP's. The annihilation
rate increases until at some time, called the equilibration time,
 the annihilation and capture rates become
equal (except for a factor of 2). For the Earth, the equilibration time is
longer than the present life time, and we could expect
``overcompensation''. In the Sun, instead,
equilibrium has been reached and $\Gamma_a \simeq C/2 \simeq constant$ in
the presence of compensation. In \cite{duda} it was shown that up-going
muon rates in underwater or under-ice km$^3$ detectors (such as IceCube,
ANTARES, NESTOR) from annihilations in the Sun (see Fig.~3) and
 the Earth remain observable even for neutralinos
constituting only 1\% of the halo dark matter.
{\epsfysize 7cm \epsfbox{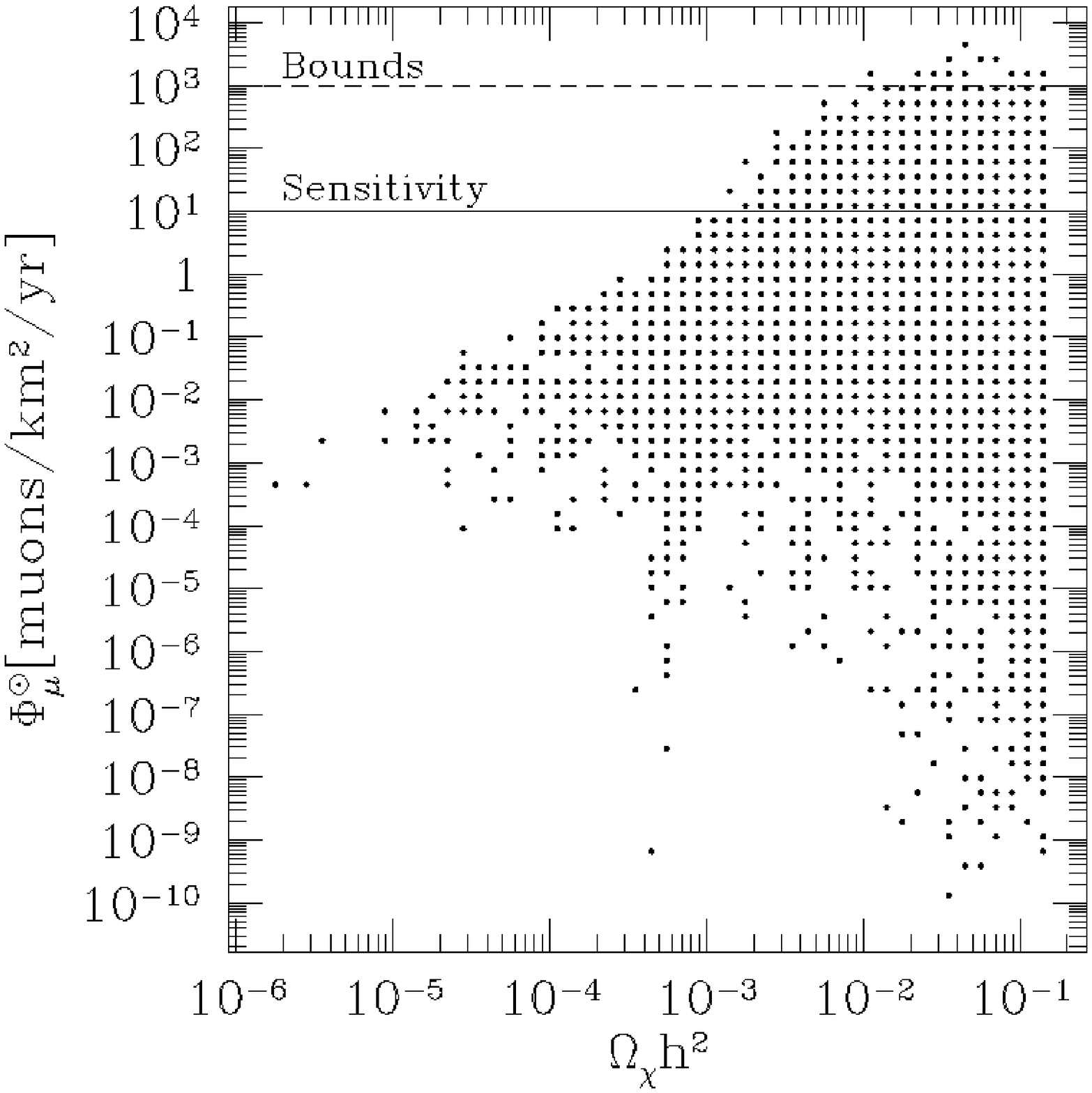}}
Figure 3. Muon fluxes in  km$^3$ detectors 
from annihilations in the Sun. The energy threshold for
detection  and the minimum observable flux  were taken to be 25 GeV
and 10 muons/km$^2$/yr.

\vspace{0.2cm}
 
``Overcompensation'' was found in neutrino signals from the Earth for
light neutralinos (m$_\chi <$ 150 GeV,) for which  signals
are largest around a halo fraction of 1\%, Except in this case, in
\cite{duda} it was not found that
the detectability of subdominant relic neutralinos is
usually favored, as was previously
stated~\cite{bottino2341}. Moreover, notice that while in \cite{duda} the value of
$\Omega_{CDM} h^2$ is fixed,  in ~\cite{bottino2341} a range of possible 
$\Omega_{CDM} h^2$ values is considered, such that as long as  $\Omega_\chi h^2$ 
is within that range
neutralinos are taken to account for the whole halo. In
  this case as $\Omega_\chi h^2$ decreases within this allowed range, detectability
increases when $\sigma_s\sigma_a^{-1} \simeq constant$, 
as cross sections become larger at constant halo density.

Many neutralino models constituting from 1\% to 100\% of the dark matter halo
are detectable in three ways (direct, indirect from the Sun, indirect from the
Earth), some in two, and some in only one. All models in \cite{duda}
detectable through annihilations in the Earth are also visible through
annihilations in the Sun (but the converse is not true) and will be either
detected or rejected already by CRESST-II. Models detectable through a
neutrino signal from the Sun may or not be seen through direct detection 
(see Fig.~4).
{\epsfysize 7cm \epsfbox{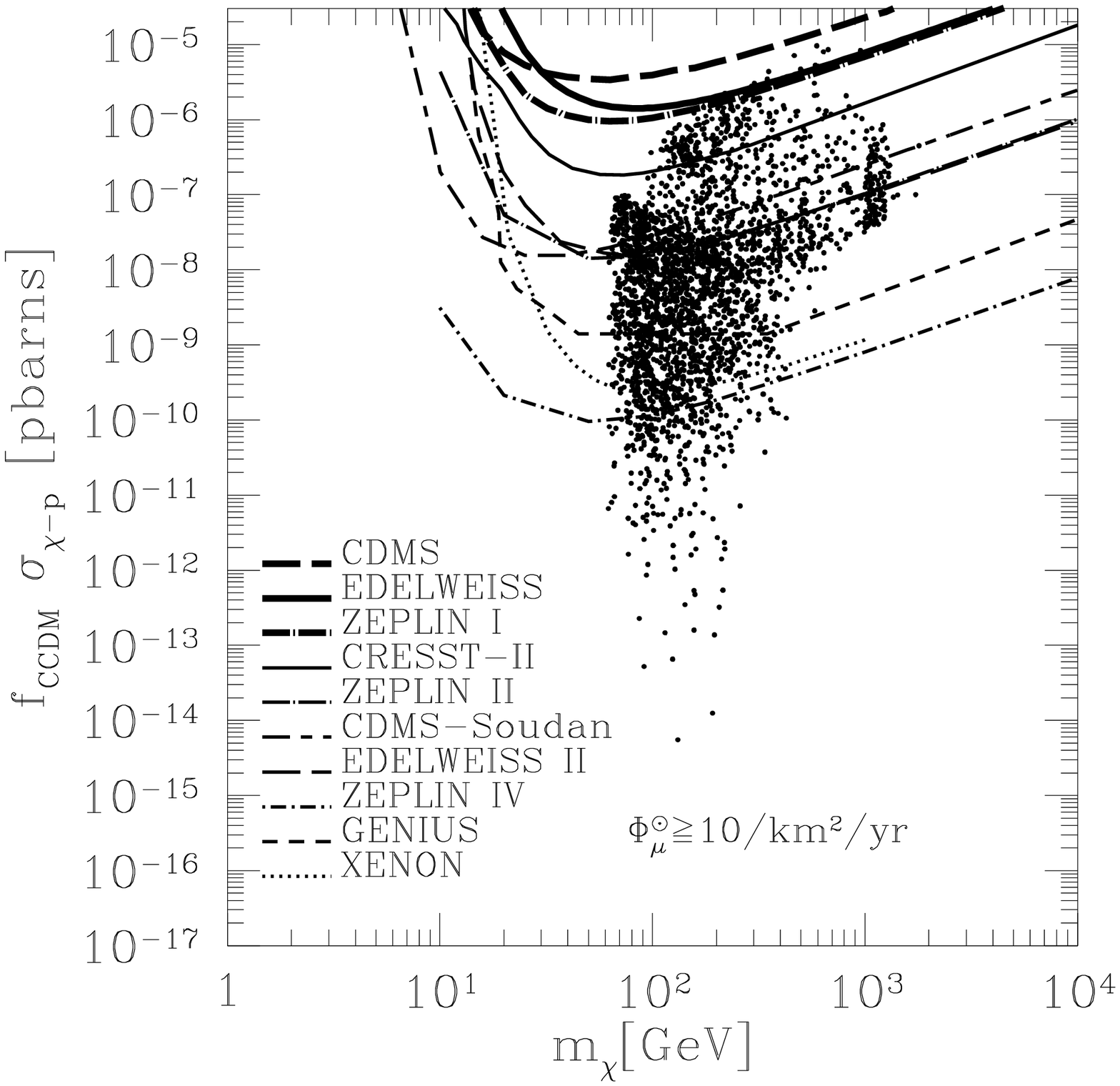}}
Figure 4.  Each point represents a model detectable through
annihilations in the Sun.

\end{document}